# An Enhanced Online Certificate Status Protocol for Public Key Infrastructure with Smart Grid and Energy Storage System

Hong-Sheng Huang[1], Cheng-Che Chuang[1,†], Jhih-Zen Shih[1,‡], Hsuan-Tung Chen[2], Hung-Min Sun[3,*]

[1, 1,†]Institute of Information Security, National Tsing Hua University, Hsinchu, Taiwan
[1]E-mail address: ray.h.s.huang@m111.nthu.edu.tw
[1,†]E-mail address: richard.chuang@gapp.nthu.edu.tw

[2]Information and Communications Research Laboratories, Industrial Technology Research Institute, Hsinchu, Taiwan
E-mail address: hsuantung@itri.org.tw

[1,‡, 3,*]Department of Computer Science, National Tsing Hua University, Hsinchu, Taiwan
[1,‡]E-mail address: s112062587@m112.nthu.edu.tw
[3,*]E-mail address: hmsun@cs.nthu.edu.tw

## Abstract

The efficiency of checking certificate status is one of the key indicators in the public key infrastructure (PKI). This prompted researchers to design the Online Certificate Status Protocol (OCSP) standard, defined in RFC 6960, to guide developers in implementing OCSP components. However, as the environment increasingly relies on PKI for identity authentication, it is essential to protect the communication between clients and servers from rogue elements. This can be achieved by using SSL/TLS techniques to establish a secure channel, allowing Certificate Authorities (CAs) to safely transfer certificate status information. In this work, we introduce the OCSP Stapling approach to optimize OCSP query costs in our smart grid environment. This approach reduces the number of queries from the Device Language Message Specification (DLMS) server to the OCSP server. Our experimental results show that OCSP stapling increases both efficiency and security, creating a more robust architecture for the smart grid.

## 1. Introduction

The Public Key Infrastructure (PKI) is a security mechanism that provides the identity of components in the environment. The major method used is the digital signature, which labels the role of every object. In the regular PKI architecture, the Certificate Authority (CA) signs the digital certificate and issues it to the applicant. Generally, the applicant is the Registration Authority (RA), which accepts the client's certificate signing request (CSR) to apply for a valid certificate. The RA checks the content of the CSR, and if the client's application is approved, it transfers the CSR to the CA for signing. The CA then generates a valid certificate for the client and sends it back to the RA. The RA checks the certificate. This process follows the most popular standard for Internet Public Key Infrastructure, X.509[1].

Building on this concept, Hsu et al. [2] constructed the X.509v3 architecture for the smart grid environment. In their work, the client became the meter manufacturer, which applies the CSR to the RA using the SCEP and EST protocols. These protocols are standard RFC 8894 [3] and RFC 7030 [4], proposed by the IETF. They also proposed a certificate whitelist authentication method shown in Figure 1. In this method, a trusted meter manufacturer can directly obtain temporary certificates for the meter. Meters can then apply for an enrollment request to the Root CA Server for signing. However, if the meter manufacturer is untrusted or if they need to update the certificates, they should follow the procedure outlined in Figure 2. This involves sending the request to the RA server for checking the content of the CSR, then passing the request to the CA server.

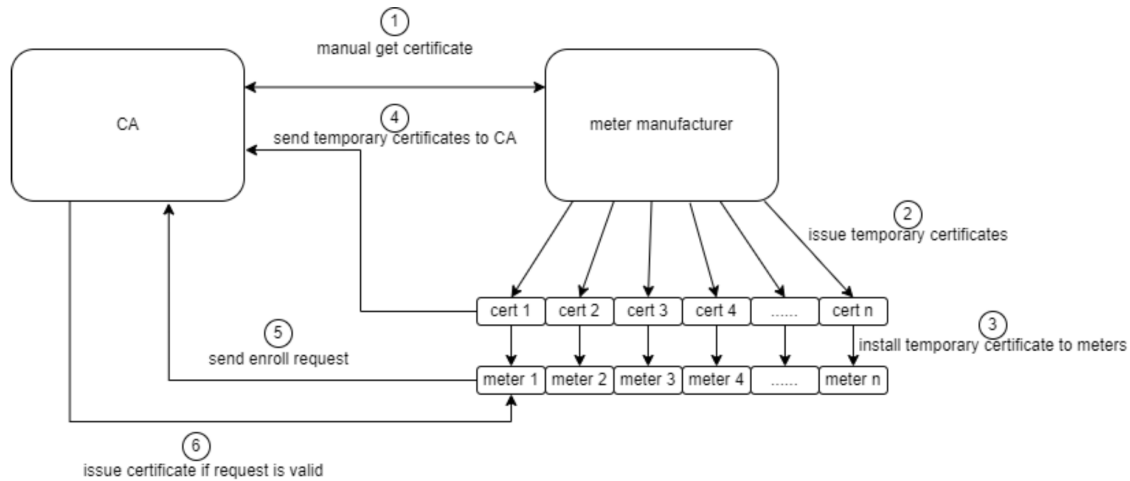

Figure 1. Certificate whitelist authentication method [2]

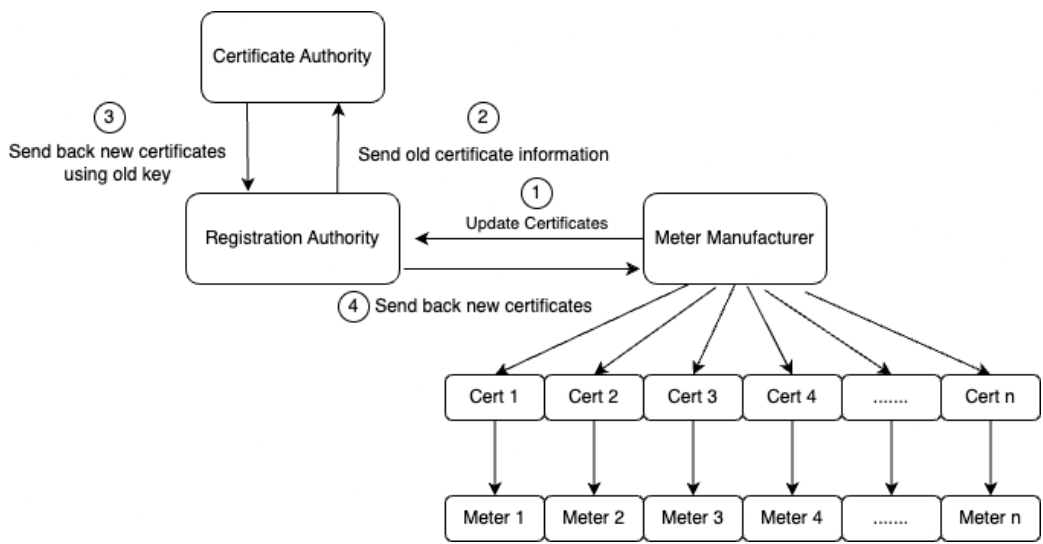

Figure 2. Certificate Update Procedure [5]

In addition to the robustness of certificate generation, auditing the validation of certificates is also a significant task. The X.509 standard introduced the Certificate Revocation List (CRL), which stores information about revoked certificates. Typically, the CA records the certificate serial number, revocation time, and revocation reason, as shown in Figure 3. Clients can check the status of a specific certificate from the download link with their enrolled certificate, as shown in Figure 4. The CRL is public, allowing every developer or other system components to view the revoked certificates. However, all revoked items are stored in one list, which grows as the system continues to operate. This growth increases the cost of downloading and analyzing the list. To address this problem, the IETF proposed the RFC 6960 [5] standard, known as the Online Certificate Status Protocol (OCSP).

The OCSP can store the certificate status along with the CA server. When the client wants to check the validity of a certificate, it queries the OCSP, which returns the result in a typical architecture. The standard version handles all query requests, as shown in Figure 5(a). This places a significant computational load on the CA server. To address this, Hsu's study restructured the architecture, making the OCSP an independent server to handle query requests from clients, as shown in Figure 5(b). This reduces the depletion of computational resources. However, in Hsu et al.'s work, they only considered an independent OCSP server to handle all query requests from clients. In the smart grid environment, the client could be a smart meter or meter manufacturer, typically involving tens of thousands of endpoints. Therefore, relying on a single OCSP server is insufficient.

Figure 3. Certificate Revocation List [6]

Figure 4. CRL Download Link [6]

Huang et al. [6] proposed the Hybrid OCSP server to address this concern. This server utilizes the CRL, which stores revoked certificate information, and integrates it with the OCSP server to download this information. Their Hybrid OCSP framework incorporates both a whitelist and a blacklist design. The blacklist is updated every hour to ensure currently status. When a client sends a query request, the OCSP server searches the CRL. If the requested certificate is in the CRL, the server responds that the certificate is invalid. If the certificate is not in the CRL, it remains valid. By following this procedure, the OCSP server does not need to search the entire certificate database for each query request, which decreases the load on the OCSP server.

Furthermore, Huang et al. split the architecture of the Hybrid OCSP framework into distributed OCSP servers, as shown in Figure 6. Each Hybrid OCSP server, called a Proxy OCSP server, handles a certain number of smart meters. These Proxy OCSP servers are maintained by specific smart manufacturers in the smart grid environment, while the CRL is maintained by the CA server. Even if an OCSP server goes down temporarily, it can quickly warm up and recover the revoked certificate data from the CA server. Their design ensures uninterrupted certificate validation and achieves high availability.

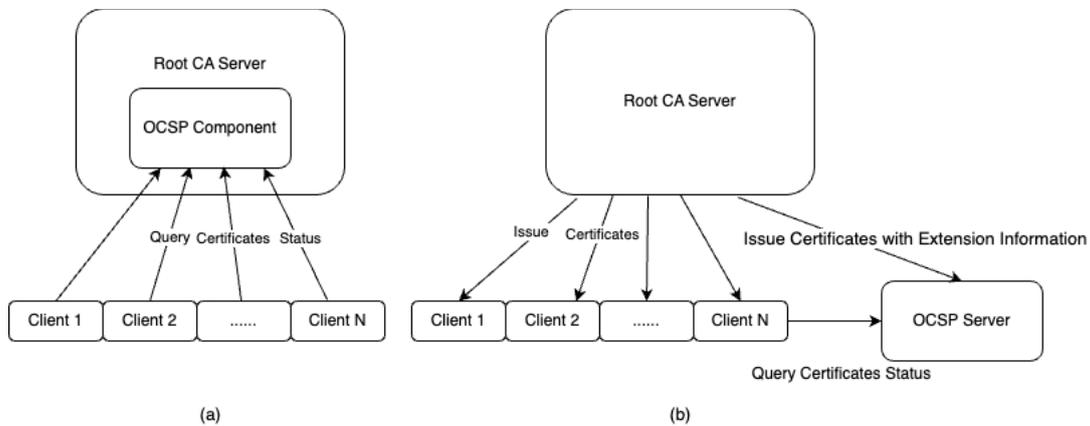

Figure 5. (a) The standard version OCSP along with the CA Server. (b) The Independent OCSP Server [2]

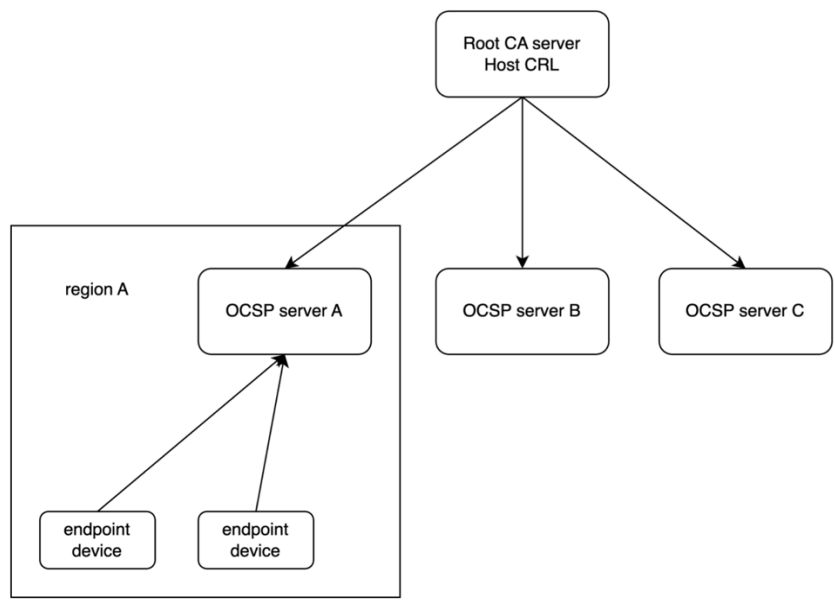

Figure 6. Hybrid OCSP with Proxy OCSP Server

However, in the smart grid scenario, there are usually tens of thousands of smart meters. Even considering the distributed OCSP server, where each server oversees its endpoint devices, the number may still be a few thousand. When a client wants to check the status of a certificate, they need to initiate an additional request each time, increasing the query cost. To improve query efficiency, we propose utilizing OCSP stapling to optimize Huang's work. This method binds the OCSP response to the smart meter certificate request and returns it along with the OCSP response to the client. This approach saves time and resources and provides a safer mechanism for PKI in the smart grid scenario.

## 2. Related Work

Although the design of PKI allows every single object to directly trust all elements, such as in Hsu's study, which mentions that the client and Root CA have a whitelist that can immediately update the certificate without checking the client's identity when the client has contacted the RA, sophisticated cyber-attackers still pose a threat to the security of the public key infrastructure. This issue is discussed in "CCSP: A Compressed Certificate Status Protocol" proposed by Chariton et al. [7] and "Design and Implementation of a Compressed Certificate Status Protocol" proposed by Pachilakis et al. [8]. Their work points out that attackers may trick web browsers into trusting a revoked certificate as a valid one.

The concrete threat in their work assumes that the attacker can launch a man-in-the-middle (MITM) attack by controlling web servers. This situation can occur in scenarios such as the attacker controlling public and free Wi-Fi, or controlling the VPN the victim uses. Moreover, setting up a rogue Wi-Fi router, like in an Evil Twin attack, can deceive the victim into connecting to the access point. In this case, the adversary must be a legitimate element in the system, such as a valid owner of a certificate, and forge the response from the DNS or HTTP endpoint to successfully impersonate the client of the certificate.

Subsequently, the adversary could obtain secret information if the victim wrongly trusts the impersonated legitimate party. In Chariton's study, they aim to make the user aware of the forgery attack when connecting to the attacker and to prevent the leakage of private information. Based on these factors, a possible attack scenario could occur when the victim falls into the pitfall of connecting to a fake website with a revoked certificate. The attacker, able to launch a MITM attack, presents the revoked certificate to the victim. This revoked certificate was valid in the recent past, so it is referred to as an "old" certificate.

The attacker tries to convince the victim that the certificate has not been revoked. The adversary would utilize a replay attack to replay an old certificate with a signed OCSP stapling response, claiming the certificate is still valid. Since the environment is controlled by the attacker, the attacker provides the old certificate, and the victim uses this certificate to visit the malicious website created by the attacker. The old certificate would appear valid to the victim when communicating with the fake website.

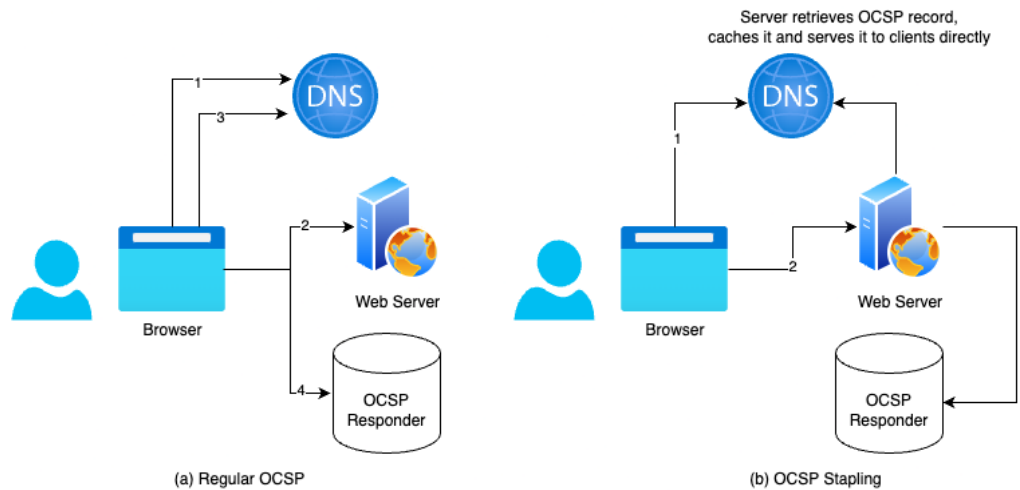

Figure 7. OCSP Stapling [9]

Another man-in-the-middle case occurs when OCSP responses are served by a Content Delivery Network (CDN), as shown in figure 7. As mentioned before, the attacker controls the entire network scenario, allowing them to impersonate both the website and the CDN-based OCSP responder. When the victim connects to the website using their own browser, the website will show the old certificate with a signed OCSP response as valid, making the client believe their certificate is still valid. When the victim falls into this illusion, they might submit their real password to the website or enter personal private information, depending on the website's functionality. At this point, the attacker can steal the legitimate identity, impersonate the victim, or trick them into installing malicious software via the browser.

This kind of attack may be considered rare and difficult to implement, but the potential damage is significant. A MITM attack could lead to loss of currency, leakage of personal information, and, in critical systems such as medical devices or national critical infrastructure (including nuclear power plants, refineries, and smart grids), severe consequences. The penalty loss depends on the significance of the facility. In a PKI environment, the victim cannot typically distinguish whether a website is fake, as the system is regarded as an intranet. If the attacker has the highest permissions, such as the role of Root CA, they can create all the bogus information for the client.

Chariton's work mentions studies that have measured the revocation rate at around 8% [10, 11], or 180,000 in absolute numbers, as measured by the popular Certificate Authority "Let's Encrypt." This authority uses an automatic program called "Certbot" to update the CRL at specific intervals, typically every seven days. In some cases, during this window of vulnerability, attackers may attempt to exploit vulnerabilities like Heartbleed against the revocations.

To defend against this kind of attack, one way to penetrate the fake environment is to introduce a timely information mechanism. This would require an additional timestamp with the current time, signed by a trusted third party. In this situation, OCSP stapling can meet these demands, as the CA would frequently check timestamps and sign all OCSP stapling responses. When a client uses an old certificate to communicate with a website, the response would be directly rejected, and the revocation information would be prompted.

Obviously, very frequently checking the timestamp and signing all OCSP stapling responses would place a tremendous burden on the CA and OCSP responder in the general architecture, requiring them to timestamp, sign, and distribute millions of responses per second. Chariton et al. addressed this overhead by introducing the "Signed Collections" method, which is an abstraction that packs revocation information not for a single certificate, but for a collection of certificates in a single response. This approach significantly reduces the number of required signatures by several orders of magnitude. In their work, they demonstrate that 1,000,000 revocation bits can be packed together, requiring only one signature operation, instead of one signature operation for each certificate.

In the CCSP protocol, the revocation information for several certificates is packed into a single OCSP response, represented as a Signed Collection (SC), which is a bitmap. Each bit corresponds to the revocation status of a single certificate: the certificate is revoked if the bit is "1"; otherwise, the certificate is valid if the bit is "0." These bits are called Revocation Bits. The name of the Signed Collection is determined as the certificate index in the collection table, which corresponds to the revocation status of the certificate. When a client connects to a website, it receives certificate C, which contains the name of the Signed Collection and the index I of the certificate within the collection, along with the OCSP stapling response. If the web server does not support OCSP stapling, the browser will request the response from an OCSP Responder maintained by the CDN in the system. Finally, the user can check the certificate status from the validity bit $SC[i]$: the certificate is revoked if $SC[i]$ is "1"; otherwise, the certificate is valid if the value is "0".

## 3. Method

OCSP stapling is a Transport Layer Security (TLS) status extension that allows clients to specify and support several certificate status methods, providing enhanced safety for the OCSP feature proposed by RFC 6961 [12]. It strengthens the OCSP server by providing status information about not only the server's own certificate but also the status of intermediate certificates in the chain. Clients can use this extension to request the CA's current status of its certificate. This approach has several advantages: it reduces the number of roundtrips and prevents network issues when clients verify the status of the CA's certificate. Additionally, it decreases the burden on the CA's status response server and increases efficiency when clients frequently send query requests.

In Huang's work, there are some problems with their Hybrid OCSP method. First, their OCSP server does not provide functionality to request the status information about intermediate CA certificates. This means the client must request status information via another method, such as downloading the CRL, to check the CA certificate status, which requires additional query computation. The second problem is that the current format of the extension and requirements in the TLS protocol prevent a client from offering the server multiple status methods.

Nowadays, CAs issue their intermediate CA certificates that not only specify the publication point for their CRLs in a CRL Distribution Point but also provide a URL for their OCSP server in the Authority Information Access extension. Since client-cached CRLs are frequently out of date, clients need to rely on OCSP to access up-to-date status information about intermediate CA certificates. Through this mechanism, clients benefit from TLS servers providing certificate status information, not only for the server certificate but also for the intermediate CA certificate, regardless of type.

The benefit of combining the status checks into one extension will reduce the time needed for the client to complete the handshake to just those required for negotiating the TLS connection. Additionally, for the CA, it can decrease the load on the number of certificates it has issued, rather than on the number of visitors to the website. Thus, using the OCSP stapling extension significantly reduces privacy concerns regarding clients informing the certificate issuer about which website they are visiting. We would like to introduce this multiple-OCSP mode in Huang's OCSP architecture.

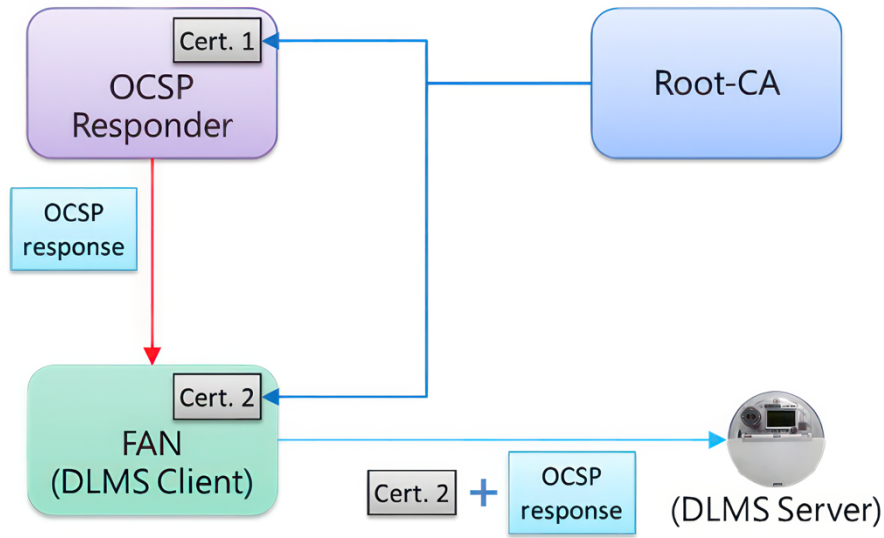

Figure 8. DLMS protocol with OCSP Stapling mechanism

In this work, we use OCSP stapling to enhance Huang's certificate revocation mechanism for optimizing our smart meter architecture as shown in Figure 8. The Device Language Message Specification (DLMS) is an international standard (IEC 62056 [13]) proposed by the DLMS User Association (DLMS UA) for electricity metering data exchange. In Huang's work, the client needs to initiate requests to their Hybrid OCSP server to obtain certificate status; each check for a single certificate requires a separate query. We aim to use OCSP stapling to combine the OCSP response and Root-CA certificate return for the DLMS client.

Thus, when our DLMS client wants to communicate with the DLMS server, the server will check the client's certificate for revocation. However, the server no longer needs to ask the OCSP server about the client's certificate status; it only needs to check the OCSP response and root CA evidence provided by the client. This reduces the round trips between the DLMS server and the OCSP server. The DLMS client will unconditionally trust the DLMS server, so before contacting the server, it sends a query request to the OCSP Responder to obtain the result.

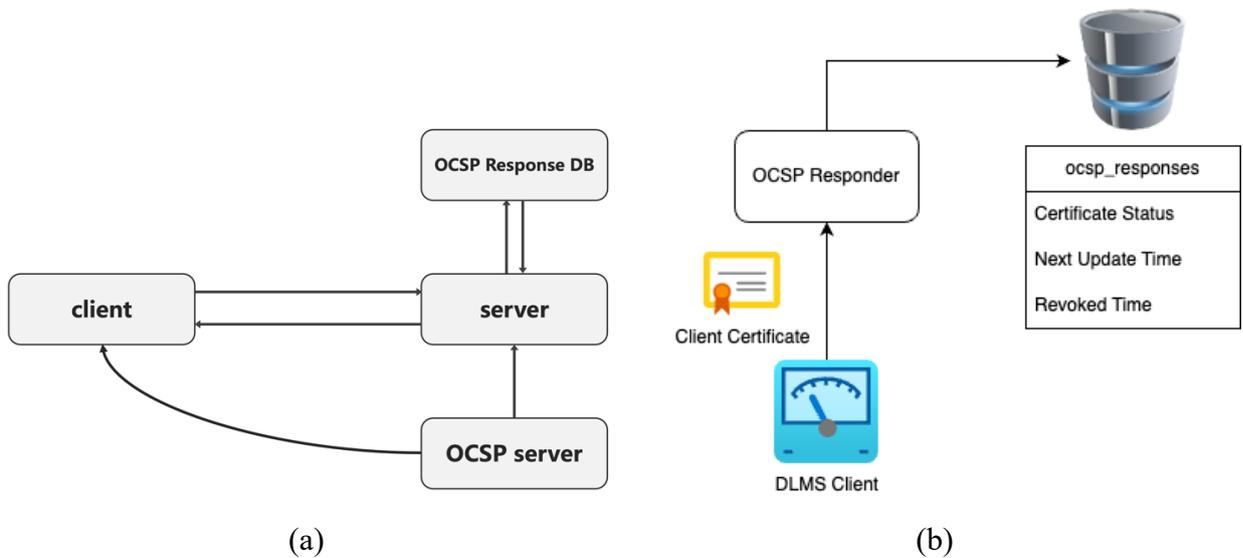

(a)          (b)

Figure 9. (a) Traditionally, the client commutes with the server that needs to check the client's identity each time. (b) In our scenario, the OCSP responder tries to save costs.

The details are as follows: We are building the OCSP Responder to handle DLMS client queries, as shown in Figure 9. The procedure is as follows: (i) Our responder analyzes the specific certificate and issuer certificate, then (ii) obtains the OCSP URL from the certificate along with the serial number of the specific certificate and issuer certificate. After requesting the response information, we use it as the identity tag for this OCSP response. (iii) To store OCSP response information, we build a database and create a table called 'ocsp_response', which can also be used to store related data for maintenance or updates, as shown in Figure 10. (iv) To exclude existing OCSP responses, each database write operation checks whether the serial number already exists in the table. (v) If the responder parses a certificate containing an OCSP URL, it establishes a request to that specific OCSP server address. The OCSP responder sends the query request to the Proxy OCSP server and retrieves the response. The responder reads and stores the following results in its own database: (1) certificate status, (2) next update time, and (3) DER format contents of the OCSP response. If the OCSP response is successful, we then extract the information from the OCSP response to the database table as shown in Figure 11.

```
sql_create_table = """
CREATE TABLE IF NOT EXISTS ocsp_responses (
    id INTEGER PRIMARY KEY AUTOINCREMENT,
    serial_number TEXT,
    cert_status TEXT,
    ocsp_url TEXT,
    next_update TEXT,
    ocsp_response BLOB,
    certificate BLOB,
    issur_certificate BLOB
)
"""
```

Figure 10. ocsp_responses table

```
ID: 2
serial_number: 120503894287240745971437000236030918190
cert_status: OCSPCertStatus.GOOD
ocsp_url: http://ocsp.pki.goog/gts1c3
next_update: 2024-06-26 09:00:42
ocsp_response: b'0\x82\x01\xd3\n\x01\x00\xa0\x82\x01\xcc0\x82\x01\xc8\x06\t+\x
90\x82\x01\xb50\x81\x9e\xa2\x16\x04\x14\x8at\x7f\xaf\x85\xcd\xee\x95\xcd=\x9c\
0043Z0s0q0I0\t\x06\x05+\x0e\x03\x02\x1a\x05\x00\x04\x14\xc7.y\x8a\xdd\xffa4\xb
x8at\x7f\xaf\x85\xcd\xee\x95\xcd=\x9c\xd0\xe2F\x14\xf3q5\x1d\'\x02\x10Z\xa84\x
00\x18\x0f20240619100043Z\xa0\x11\x18\x0f20240626090042Z0\r\x06\t*\x86H\x86\xf
0\x17BrE\x91\xc872\x16\xf5\xc1\xc1q\xe1\xdeP\xe5q|\x8cw8$\x03)\xb9~X,\xd0\xce\
-g\xeau\xffc\x12\xd0\xdc\xf7\x8d\t7h]Z\xc7u9\x97\xbd\xf5\xeb\xcd\xa6~\xaaZB\xe
\x12\xce\xc2s)\xca\xc9\toN\x82M\xef\x97\xd7\x9c\xd9\xad\\:@\x90+\xe0\xe5\xdf\x
\x97\x81?x\xa9!d\x1a&\xde"\x03\n\x97\xfd;\xbc\x07qo\xedV\x86\xa6\x8e4FQ\xbd4\x
?\xbe\x8e\xc1\xc2\x03\xc4\xd2\xd1o\xe4\xa7|\xd3td\xdd\xf2R\xe5\x00\x13\xbc\xfb
fV\xad*\xa6]\x8a^\\P:\x9b\x17\xcbr>D\xc3\xe1\xb1E\xda\xb2\xda\xa0\r$\xc4\x04\x
certificate: b'-----BEGIN CERTIFICATE-----\r\nMIIOfDCCDWSgAwIBAgIQWqg0shTZF68J
YDVQQGEwJVUzEiMCAGA1UEChMZR29vZ2xlIFRydXN0IFNlcnZpY2VzIExM\r\nQzETMBEGA1UEAxMK
A4MTMw\r\nNTMyMDJaMBcxFTATBgNVBAMMDCouZ29vZ2xlLmNvbTBZMBMGByqGSM49AgEGCCqG\r\n
```

Figure 11. The ocsp_response content in the database

To maintain the validity of OCSP responses, we have also built an automatic program that checks the database table daily, as shown in Figure 12. For each certificate, the program compares the 'next_update' time with the current time. If the 'next_update' time is less than seven days from the current time, indicating that the revocation list has a new version and needs updating, the program sends an OCSP request to update the OCSP response. Otherwise, if the difference is greater than one day, no action is required.

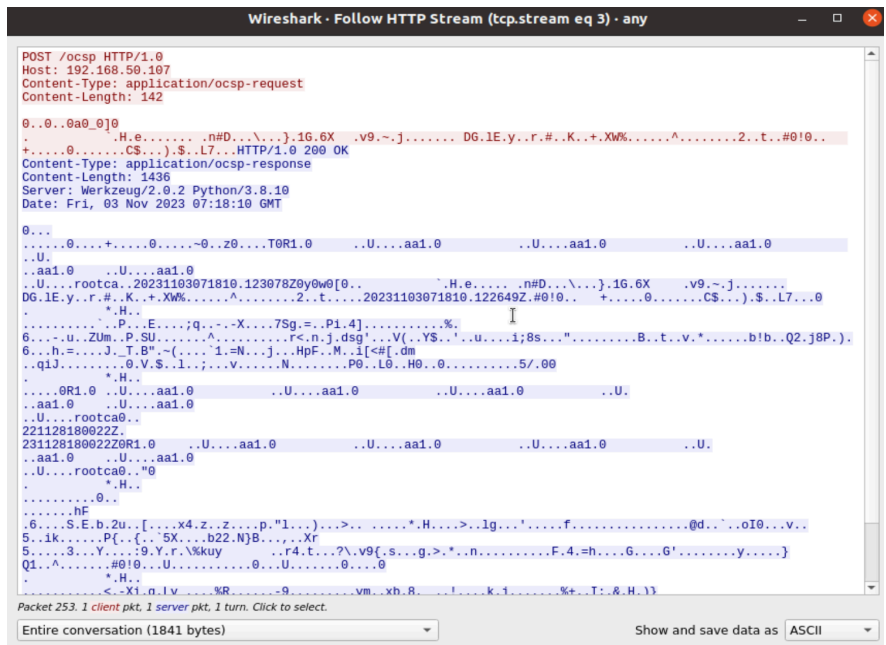

Figure 12. Maintain the ocsp_response

## 4. Result

In our experiment, the device specifications include the host operating system, Proxmox Virtual Environment, equipped with an Intel 10980EX CPU, and the guest operating system, Ubuntu 20.04.3 LTS, with 11GB of RAM, similar to Huang's settings. We utilize this virtual machine to simulate the DLMS client and measure its data consumption associated with OCSP requests via OpenSSL. The aggregated data, including the request and response, amounts to 1,841 bytes, as observed in Figure 13 using WireShark.

Figure 13. OCSP data frame [6]

In their benchmarking analysis of the Hybrid OCSP mechanism, they showed the elapsed time values and subsequently computed the average request time to handle the 1,841-byte data in Table 1. The experiment was based on the Root CA and Hybrid OCSP server developed using the Python programming language with the Flask framework, which is suitable for OCSP stapling situations. Therefore, we aim to extend their architecture by building an independent OCSP Responder server in the system.

|  | Time consumed | Avg. request time |
|---|---|---|
| 100 thousand OCSP requests on 4 cores CPU | approximate 17 mins | 0.0102s /per request |
| 1 thousand OCSP requests on 2 cores CPU | approximate 30 secs | 0.029s /per request |

Table 1. Hybrid OCSP server benchmark [6]

In this work, we successfully constructed the DLMS server and simulated DLMS clients. The DLMS server can parse the OCSP response from the client, and the OCSP Responder can handle OCSP requests and store response information in the database. To maintain the correctness of OCSP responses, we have also developed an automatic program to check the table and exclude incorrect data from the database. Thus, our mechanism reduces the OCSP query response time between the DLMS server and the Proxy OCSP server.

## 5. Conclusion

In this paper, we introduced an optimized Online Certificate Status Protocol (OCSP) using the OCSP Stapling approach and implemented it in our smart grid scenario. Based on the RFC 6961 standard, this method improves the Hybrid OCSP mechanism. Utilizing the stapling approach can resist man-in-the-middle attacks with much better robustness than previously proposed mechanisms. Additionally, using the stapling extension can reduce the communication time between the DLMS server and the distributed OCSP server, as well as save network traffic bandwidth in our case.

We successfully implemented an independent OCSP Responder to handle the DLMS client's certificate and their OCSP request. The responder can then apply the OCSP request to an upper-layer OCSP server according to the client's OCSP URL. To store the OCSP response, we created a database with an 'ocsp_response' table. We enhanced Huang's work by providing more scalability features for smart meters within the smart grid scenario. We believe that this design can provide greater safety and efficiency in the public key infrastructure.

## 6. Acknowledgements

This experimental research was supported by Dr. Hsuan-Tung Chen, who provided insights and expertise that greatly assisted the research. We acknowledge the support from the Energy Administration, Ministry of Economic Affairs, R.O.C. We would also like to express our gratitude to Professor Hung-Min Sun for sharing his pearls of wisdom with us.

## 7. References

[1] Sharon Boeyen, Stefan Santesson, Tim Polk, Russ Housley, Stephen Farrell, David Cooper. (2008). Internet X.509 Public Key Infrastructure Certificate and Certificate Revocation List (CRL) Profile. RFC 5280, IETF. https://datatracker.ietf.org/doc/html/rfc5280

[2] Hsu, DW., Chen, HT., Sun, HM., Huang, TY. (2023). ECDSA Certificate Enrollment and Authentication for SCEP Protocol in Smart Grid PKI. In: Weng, S., Shieh, CS., Tsihrintzis, G.A. (eds) Advances in Intelligent Information Hiding and Multimedia Signal Processing. IIHMSP 2022. Smart Innovation, Systems and Technologies, vol 341. Springer, Singapore. https://doi.org/10.1007/978-981-99-0605-5_26.

[3] Peter Gutmann. (2020). Simple Certificate Enrolment Protocol. RFC 8894, IETF. https://datatracker.ietf.org/doc/html/rfc8894.

[4] Max Pritikin., Peter E. Yee., Dan Harkins. (2013). Enrollment over Secure Transport. RFC 7030, IETF. https://datatracker.ietf.org/doc/html/rfc7030#section-3.2.3.

[5] Stefan Santesson., Michael Myers., Rich Ankney., Ambarish Malpani., Slava Galperin.,Carlisle Adams.(2013). X.509 Internet Public Key Infrastructure Online Certificate Status Protocol - OCSP. https://www.rfc-editor.org/rfc/rfc6960.html.

[6] Huang, H., Jiang, Z., Chen, H., & Sun, H. (2024). Hybrid Online Certificate Status Protocol with Certificate Revocation List for Smart Grid Public Key Infrastructure. ArXiv, abs/2401.10787.

[7] A. A. Chariton, E. Degkleri, P. Papadopoulos, P. Ilia and E. P. Markatos, "CCSP: A compressed certificate status protocol," IEEE INFOCOM 2017 - IEEE Conference on Computer Communications, Atlanta, GA, USA, 2017, pp. 1-9, doi: 10.1109/INFOCOM.2017.8057065.

[8] Michalis Pachilakis, Antonios A. Chariton, Panagiotis Papadopoulos, Panagiotis Ilia, Eirini Degkleri, and Evangelos P. Markatos. 2020. Design and Implementation of a Compressed Certificate Status Protocol. ACM Trans. Internet Technol. 20, 4, Article 34 (November 2020), 25 pages. https://doi.org/10.1145/3392096

[9] Igor Cicimov. (2015). HAProxy OCSP stapling. https://icicimov.github.io/blog/server/HAProxy-OCSP-stapling/

[10] Roger A. Grimes. [n.d.]. The sorry state of certificate revocation. Retrieved from https://www.csoonline.com/article/3000574/security/the-sorry-state-of-certificate-revocation.html.Google Scholar

[11] Yabing Liu, Will Tome, Liang Zhang, David Choffnes, Dave Levin, Bruce Maggs, Alan Mislove, Aaron Schulman, and Christo Wilson. [n.d.]. An end-to-end measurement of certificate revocation in the Web's PKI. In Proceedings of the ACM Internet Measurement Conference. 14. DOI:https://doi.org/10.1145/2815675.2815685Google Scholar

[12] Yngve N. Pettersen. (2013). The Transport Layer Security (TLS) Multiple Certificate Status Request Extension. RFC 6961, IETF. https://datatracker.ietf.org/doc/html/rfc6961

[13] Wikipedia. (2023). IEC 62056. DLSM User Association. https://en.wikipedia.org/wiki/IEC_62056